\documentclass[prl,twocolumn,showpacs,preprintnumbers]{revtex4}
\usepackage{graphicx}

\begin{document}

\preprint{hep-ph/0607147}
\preprint{MIFP-06-18}

\title{$B_s$-$\bar B_s$ Mixing in Supersymmetric Grand Unified Models
}


\author{Bhaskar Dutta}
%\thanks{}
\author{Yukihiro Mimura}
%\email[]{mimura@physics.tamu.edu}
%\thanks{}
\affiliation{
Department of Physics, Texas A\&M University,
College Station, TX 77843-4242, USA}


\date{\today}

\begin{abstract}
We study  $B_s$-$\bar B_s$ mixing in Supersymmetry grand unified SO(10), SU(5)
models where the mixings among the second and third generation
 squarks arise due to the existence of flavor violating sources in the
Dirac and Majorana couplings which are responsible for neutrino
mixings. We find that when the branching ratio of
$\tau\rightarrow\mu\gamma$ decay is enhanced to be around the
current experimental bound, $B_s$-$\bar B_s$ mixing may also contain
large contribution from supersymmetry in the SO(10) boundary
condition. Consequently, the phase of $B_s$-$\bar B_s$ mixing is
large (especially for small $\tan\beta$ and large scalar mass $m_0$)
and can be tested by measuring CP asymmetries of  $B_s$ decay modes.
\end{abstract}

%\keywords{}

\maketitle

%\section{}
%\subsection{}
%\subsubsection{}

%Recent measurement of $B_s$-$\bar B_s$ mixing D$\O$ and CDF
%collaboration \cite{Abazov:2006dm}
%is important
%not only to test Kobayashi-Maskawa theory \cite{Kobayashi:1973fv}
%but also to probe new physics.

Flavor changing processes are important not only to test the
Kobayashi-Maskawa theory \cite{Kobayashi:1973fv} and to determine
the parameters in CKM (Cabibbo-Kobayashi-Maskawa) matrix, but also
to examine new physics.
Recent measurement of $B_s$-$\bar B_s$ mass difference,
\begin{equation}
\Delta M_s = 17.77 \pm 0.10 \pm 0.07 \ {\rm ps}^{-1},
\end{equation}
by D\O\ and CDF collaborations~\cite{Abazov:2006dm}
can impact on the flavor structure of new physics beyond the
standard model (SM)~\cite{Goto:2003iu,Endo:2006dm}. %$\Delta M_s$
%
%The mass differnce of $B_s$-$\bar B_s$ is
%
%
%\begin{equation}
%\Delta M_s = 17.31^{+0.33}_{-0.18}
%\pm 0.07 \ {\rm ps}^{-1}\,,
%\end{equation}
%
The experimental constraints for new physics are not very severe yet
since deviations from the SM prediction can be buried in the errors
of CKM parameters and lattice calculation. In other words, there is
still  room for new physics. However, the parameters are expected to
be determined more accurately in the near future. Besides, the CP
asymmetry of $B_s$ decay can be observed by direct measurements, and
we will get important information of new physics.

Supersymmetry (SUSY) is one of the most promising candidates of new
physics. SUSY can provide a natural prospect to have a large
hierarchy in the theories, and in the minimal SUSY standard model
(MSSM), gauge forces can unify at a high scale, which leads to a
successful realization of grand unified theories (GUTs).
However, the flavor sector has not yet been well accepted in the
MSSM due to the fact that SUSY breaking terms can induce large
flavor changing neutral currents (FCNCs).
%,
%and have  potentials to modify the flavor structure
%drastically.
%
Actually, the experimental constraints of FCNCs introduce  flavor
degeneracy of the SUSY particles, especially in first  and second
generations, if SUSY particles are lighter than around 2-3 TeV
\cite{Gabbiani:1988rb}.

In order to suppress the SUSY FCNCs, the squarks and sleptons are
assumed to be degenerate at the GUT  scale as a nature of the SUSY
breaking.
%namely
%
%
Even though the degeneracy is realized at a scale, the
renormalization group equation (RGE) flow induces flavor violation
for squarks and sleptons at low energy. In this scenario, the flavor
violating effects in SUSY breaking terms are  small at the weak
scale  and satisfy the current FCNC constraints.
The small flavor violations originate from mixings in the Yukawa
couplings characterized by the CKM mixings
%in quark sector
as well as the neutrino mixings. %in lepton sector.
In the MSSM, the induced FCNCs in the quark sector are not large
since the CKM mixings are small. In the lepton sector, on the other
hand, sizable FCNC effects can be generated and a testable amount of
flavor violating lepton decay can be obtained
\cite{Borzumati:1986qx}, %since the neutrino Dirac
%Yukawa coupling matrix can have large off-diagonal elements,
which is related to the large mixings for the neutrino oscillations.

In the grand unified models, the flavor violation at the weak scale
can be related to the GUT scale physics.
Actually, as a consequence of the quark-lepton unification,
%This unification  relates the flavor violations in the quark and
%lepton sectors.
the large neutrino mixings not only introduces flavor violations in
the lepton sector, but also in the quark sector as well. The
relation of  flavor violation in the quark and the lepton sectors
depends on unification of  matters and how to obtain light neutrino
masses. Therefore, investigating the FCNC effects, we may obtain a
footprint of the GUT models.

The new result on $B_s$-$\bar B_s$ mixing can restrict the flavor
violation in the quark sector involving  $b$ and $s$ quarks. In
grand unified models, the $B_s$-$\bar B_s$ oscillation can be
correlated to the
%branching ratio of
$\tau\to\mu\gamma$ decay. Since  Br($\tau\rightarrow\mu\gamma$) is
being measured at the B factories, the future results will be able
to probe new contributions from the GUT models.  In this letter, we
calculate $B_s$-$\bar B_s$ mixing and Br($\tau\rightarrow\mu\gamma$)
in  SU(5) and
 SO(10) GUT models, and study  the implication of the correlation between $B_s$-$\bar B_s$ mixing and
 Br($\tau\rightarrow\mu\gamma$) decay in the CP asymmetry of
$B_s$-$\bar B_s$ mixing (which is under experimental investigation)
to decipher GUT models.

%In this letter, we calculate $B_s$-$\bar B_s$ mixing with the SU(5)
%and the SO(10) GUT models, and study an implication
% from the relation between $B_s$-$\bar B_s$ mixing and
%the branching ratio of $\tau\rightarrow\mu\gamma$ decay.
%
%Since the Br($\tau\rightarrow\mu\gamma$)
%is being measured at the B factories, the future results will be
%able to probe the contributions from the GUT models. We also
%calculate the CP asymmetry of $B_s$-$\bar B_s$ mixing
%(which is under experimental investigation) to decipher GUT models.

 The
existence of second-third generation (23) mixing elements in squarks
and slepton mass matrices generate $B_s$ mixing and
$\tau\to\mu\gamma$ decay respectively. We first investigate the
 arise of 23 elements in squark
and slepton mass matrices in the grand unified theories.
%The 23 elements in left- and right-handed down-type squark
%mass matrices are constrained at weak scale \cite{Endo:2006dm}.

{\bf SU(5)} model: In a SU(5) grand unified model, the
superpotential involving the Yukawa couplings is as follows:
$W_Y={Y_u}_{ij}/4\, {\bf 10}_i {\bf 10}_j  H_{\bf 5} +\sqrt2
\,{Y_d}_{ij} {\bf 10}_i {\bar {\bf 5}}_j  \bar H_{\bar {\bf 5}} +
{Y_\nu}_{ij}  {\bar {\bf 5}}_i N_j H_{\bf 5} +{M_{\nu}}_{ij}/2\,
N_iN_j$, where ${\bar {\bf 5}}$ contains the right-handed down-type
quarks ($D^c$)
 and left-handed lepton doublets ($L$), and $i,j$ denote the generation indices.
The left-handed quark doublets ($Q$), right-handed up-type quarks ($U^c$),
and right-handed charged-leptons ($E^c$) are unified in $\bf 10$ multiplet,
and $N$ is the right-handed neutrino.
Since two large neutrino mixings have been observed in nature, the
$Y_{\nu}$ coupling is expected to have large off-diagonal elements, which will
generate off-diagonal terms in the SUSY breaking scalar mass matrices
%of the right-handed down squarks and slepton doublets
for the scalar $\bar {\bf 5}$ multiplet via $\bar H_{\bar {\bf 5}}$
and $N$ loops~\cite{Borzumati:1986qx}. One, thus, expects a large 23
element in the SUSY breaking scalar mass matrices for $\tilde D^c$
and $\tilde L$ due to the large atmospheric mixing and possibly
large $\nu_\tau$ Dirac Yukawa coupling.
%
%get affected in this way since they
%couple to $N$ by $Y_\nu$ and
% their SUSY breaking mass matrices at the GUT scale
%become
%$M_{\tilde D^c}^2=M_{\tilde E^c}^2=m_0^2\,(I-\kappa \,Y_{\nu}Y^\dag_{\nu})$
%via $\bar H_{\bar {\bf 5}}$ and $N$ loops,
%neglecting GUT and Majorana neutrino mass scale thresholds,
%where
% $\kappa\simeq {1/8\pi^2(3+A_0^2/m_0^2)\ln(M_*/M_G)}$, $M_*$ is the string
% scale and $M_G$ is the GUT scale, $A_0$ is a scalar trilinear coupling coefficient
% and $m_0$ is a universal SUSY breaking mass.
% One, thus, expects a large 23 element in the
%SUSY breaking scalar mass matrices for $\tilde D^c$ and $\tilde L$
%due to the large atmospheric mixing.
%
Therefore, the SUSY contribution for
%$\Delta M_s$
the amplitude of $B_s$-$\bar B_s$ mixing
can be enhanced along with the branching ratio of $\tau
\rightarrow \mu \gamma$.
%In this paper we are only discussing the
%23 element and its effects in quark and lepton flavor violation. One
%can also have 12 and 13 mixing in the squarks and sleptons as well.
%However, they can be small by the choice of 13 mixing in $Y_\nu$ and the
%hierarchical pattern of the neutrino coupling.
%If these elements are large, $\mu \rightarrow e \gamma$ will become large.

Since our purpose is to investigate the flavor violation
 in the 23 sector of squark and sleptons
and to probe how it relates the $B_s$-$B_s$ mixing and flavor
violating $\tau$ decay, we are only discussing the 23 element and
its effects in quark and lepton flavor violation in this letter.
%
%of squarks from the CDF data and
%see its prediction in the leptonic sector and thereby
%search the footprint of the GUT models,
%
For this purpose, we consider the
following simplified SU(5) boundary
condition for the SUSY breaking scalar mass matrices at the GUT
scale where SU(5) is broken to the SM:
\begin{eqnarray}
&&M_{\bf 10}^2 = M_{\tilde Q}^2 = M_{\tilde U^c}^2 = M_{\tilde E^c}^2 =
m_0^2 \,{\bf 1},
\\
&&M_{\bar {\bf 5}}^2 = M_{\tilde D^c}^2 = M_{\tilde L}^2
= \left(\begin{array}{ccc}
1 & 0 &0 \\
0 & 1 & \delta \\
0 & \delta^* & 1
\end{array}\right) m_0^2 \,.
\end{eqnarray}
%
%\begin{eqnarray}
%&&M_{\tilde Q}^2 = M_{\tilde U^c}^2 = M_{\tilde E^c}^2 = m_0^2 \,{\bf 1},
%\\
%&&M_{\tilde D^c}^2 = M_{\tilde L}^2
%= \left(\begin{array}{ccc}
%1 & 0 &0 \\
%0 & 1 & \delta \\
%0 & \delta^* & 1
%\end{array}\right) m_0^2 \,.
%\end{eqnarray}
%
%where $\tilde Q$, $\tilde U^c$ and $\tilde D^c$ denote the
%left-handed squark, right-handed up- and down-type squarks
%respectively, and $\tilde L$, $\tilde E^c$  denote the left- and
%right-handed sleptons respectively.
The $\delta$ denotes the flavor
mixing term arising from the neutrino Yukawa couplings discussed
above.
%Note that the notation of SUSY breaking masses is
%$-{\cal L} = (M_{\tilde Q}^2){}_{ij} \tilde Q_i \tilde Q_j^\dagger
%+(M_{\tilde U^c}^2){}_{ij} \tilde U^c_i \tilde U_j^{c\dagger} +
%\cdots$.
We assume the above boundary condition in the basis where
the down-type quark Yukawa matrix is diagonal (at the GUT scale):
%
%The minimal SU(5) boundary condition is the following:
%
$
Y_d = Y_d^{\rm diag},\,Y_u= V_{\rm CKM}^{\rm T} Y_u^{\rm diag} P_u V_{uR},
% \\
Y_e = V_{eL} Y_e^{\rm diag} P_e V_{eR}^\dagger,
$
where the up- and down-type quarks and the charged-lepton
Yukawa couplings $Y_{u,d,e}^{\rm diag}$ are real
(positive) diagonal matrices and $P_{u,e}$ are diagonal phase
matrices. In a minimal SU(5) GUT, in which only $H_{\bf 5}$ and
$\bar H_{\bar {\bf 5}}$ couple to matter fields, we have $V_{uR} = V_{\rm
CKM}$, $V_{eL} = V_{eR} = {\bf 1}$, and $Y_d^{\rm diag} = Y_e^{\rm
diag}$. %We assume $V_{uR}, V_{eL}, V_{eR} \simeq \bf 1$ in the
%non-minimal SU(5) GUT.
%When we neglect colored Higgs loop and
%run the MSSM RGE running with boundary conditions mentioned above,
%$P_u V_{uR}$ and $P_e V_{eR}^\dagger$ are unphysical rotations.
We do not assume the minimal choice of Higgs fields, but assume
$V_{eL}\simeq {\bf 1}$ to keep the relation of flavor violation
between the quark and the lepton sectors.

We note that one can also have first-second  and first-third
generations  mixings in squarks and sleptons. However, they can be
small by the choice of 13 mixing in $Y_\nu$ and the hierarchical
pattern of the neutrino coupling. If these elements are large, $\mu
\rightarrow e \gamma$ will become large.

%Note that we use a basis $Y_{d,e}$ is real (positive) diagonal
% when we calculate observable at weak scale.
%In this case, the boundary condition is
%$M^2_{\tilde D_c} \simeq P_e M^2_{\tilde L} P_e^\dagger$.
%
%$P_e$ is
%Our purpose is to see the correlation
%of $B_s$-$\bar B_s$ mixing and Br($\tau \rightarrow \mu \gamma$).
%For this purpose, phase $P_e$ is not very important.
%
%
%Actually, the left-handed Majorana coupling arises  naturally
%in the left-right symmetric type of models.

{\bf SO(10) model}: In a SO(10) model, the flavor violations can be
more enhanced compared to the SU(5) case since all matters are
unified in spinor representation and couple to ${\bf 10}$ and
$\overline{\bf 126}$ Higgs fields~\cite{Babu:1992ia}, e.g., $ W_Y =
\frac12 {h}_{ij} {\bf 16}_i {\bf 16}_j  {\bf 10} + \frac12 f_{ij}
{\bf 16}_i {\bf 16}_j \overline {\bf 126}$. The mixings in the
neutrino Dirac Yukawa coupling may be small since right-handed
neutrinos are  also unified with other quarks and leptons.
%The flavor
%violations can arise just like in the SU(5) case.
In a SO(10) model, however,
there could be sources for large flavor violations in Majorana
couplings for both left- and right-handed neutrinos in the type II
seesaw scenario~\cite{Schechter:1980gr}.
The Majorana couplings are unified to the
${\bf 16} \cdot {\bf 16} \cdot {\overline{\bf 126}}$
coupling,
and
also affect the quark fields.
%${\bf 16} \cdot {\bf 16} \cdot
%{\overline{\bf 126}}$ coupling includes both left- and right-handed
%Majorana neutrino coupling which introduce flavor violation in both
%left- and right squark mass matrices below the GUT scale and
The couplings will
give rise to observable amount of flavor violations to the
sparticle mass matrices via the GUT particle loops.
%can be
%$M_{16}^2=m_0^2(I-\kappa' f_{\nu}f^{\dag}_{\nu})$.
%
%
Based on the above discussions, the following mass terms can arise
in a SO(10) model at the GUT scale:
\begin{eqnarray}
M_{\bf 16}^2 =
\left(\begin{array}{ccc}
1 & 0 &0 \\
0 & 1 & \delta \\
0 & \delta^* & 1
\end{array}\right) m_0^2.
\label{symbc}
\end{eqnarray}
%
%\begin{eqnarray}
%M_{\tilde Q}^2 = M_{\tilde U^c}^2 = M_{\tilde D^c}^2
%= M_{\tilde L}^2 = M_{\tilde E^c}^2 = \!
%\left(\begin{array}{ccc}
%1 & 0 &0 \\
%0 & 1 & \delta \\
%0 & \delta^* & 1
%\end{array}\right) m_0^2.
%\label{symbc}
%\end{eqnarray}
%
When the matters couple to only $\bf 10$ and $\overline{\bf 126}$
Higgs fields, the Yukawa matrices are symmetric and the boundary
condition is
$
Y_d= Y_d^{\rm diag} P_d,\,Y_u = V_{\rm CKM}^{\rm T} Y_u^{\rm diag} P_u V_{\rm CKM},
% \\
Y_e = V_{ql} Y_e^{\rm diag} P_e V_{ql}^{\rm T}.
$
We do not assume the minimal choice of Higgs fields but assume
$V_{ql} \simeq {\bf 1}$.
It needs to be noted  that the diagonal phase matrices $P_{u,d,e}$
can not be rotated away for this boundary condition and
these parameters enter into our calculations which we will see later.

It appears that  in order to suppress the proton decay and to obtain
the correct fit to fermion masses one needs to extend the above minimal
SO(10) model. The new superpotential includes {\bf 120} Higgs field:
$W_Y
= \frac12 h_{ij} {\bf 16}_i {\bf 16}_j {\bf 10}
+ \frac12 f_{ij} {\bf 16}_i {\bf 16}_j {\overline
{\bf 126}} + \frac12 h^\prime_{ij} {\bf 16}_i {\bf 16}_j {\bf 120}$.
In this case, the
symmetric nature  of the Yukawa matrices is lost since {\bf 120}
Higgs coupling $h^\prime$ is antisymmetric.
%Hermitian Yukawa matrices can be obtained by introducing a parity symmetry \cite{Dutta:2004hp}.
In this context, the Hermitian Yukawa matrices can be considered with $\bf 120$ Higgs
and a parity symmetry to reduce the number of parameters and to
solve SUSY CP problem~\cite{Dutta:2004hp}.
% one can have a correlation
%between $B_s$-$\bar B_s$ mixing and Br($\tau
%\rightarrow\mu\gamma$).
%The boundary conditions for Yukawa matrices are
%%
%\begin{eqnarray}
%Y_u &=& V_{\rm CKM}^{\rm T} Y_u^{\rm diag} S_u V_{\rm CKM}^*, \\
%Y_d &=& Y_d^{\rm diag} S_d, \\
%Y_e &=& V_{ql} Y_e^{\rm diag} S_e V_{ql}^{\dagger},
%\end{eqnarray}
%
%where $S_{u,d,e}$ are diagonal signature matrices.
The SO(10) symmetry is broken in the basis where $Y_d$ is diagonal,
since the left- and right-handed fields are rotated by conjugated
unitary matrices, e.g. $Q \to VQ$ and $U^c \to V^* U^c$. In the
original SO(10) basis, the SUSY breaking mass matrices are real by
the parity symmetry. As a result,  in the basis where the down-type
quark Yukawa matrix is diagonal, the squarks and slepton masses at
the GUT scale are related by the following relation
\begin{equation}
M_{\tilde Q}^2 = M_{\tilde U^c}^{2*} = M_{\tilde D^c}^{2*}
= M_{\tilde L}^2 = M_{\tilde E^c}^{2*}\,. %=
%\!
%\left(\begin{array}{ccc}
%1 & 0 &0 \\
%0 & 1 & \delta \\
%0 & \delta^* & 1
%\end{array}\right) m_0^2.
\label{herbc}
\end{equation}
So we see that two mass relations are possible in a SO(10) model at
the GUT scale. We call one of them symmetric (since the Yukawa
couplings are symmetric) and the other one hermitian (since the
Yukawa couplings are hermitian).
%
%In this case, the Yukawa matrices do not have extra physical phases,
%but the phase of 23 elements of SUSY breaking masses does not get
%canceled in $\Delta M_s$.
%In this case, the phase of $\delta$ does
%not get canceled in $\Delta M_s$. For example, when $\delta$ is real
%and $S_d = {\rm diag}(\pm,+,+)$, the gluino contribution
%(\ref{gluino-contribution}) is negative.

%Since we see that the flavor mixings arise naturally
%in the mass matrices for the slepton and the squark
% in the GUT models,
% we can expect effects in the $B_s$-$\bar B_s$ mixing
%and Br($\tau\rightarrow\mu\gamma)$.
 We use the above mass matrices for the boundary condition
at the GUT scale and then calculate the masses at the weak scale by
using
 RGEs
to calculate the mixing of $B_s$-$\bar B_s$
 and  Br($\tau \to \mu\gamma$).
In calculating the mass differences of mesons, one
encounters non-perturbative factors originating from
strong interaction. In the ratio of $B_s$ and $B_d$
mass differences, many common factors cancel, and the
ratio can be calculated more accurately rather than
the respective mass differences. We have
\begin{equation}
\frac{\Delta M_s^{\rm SM}}{\Delta M_d^{\rm SM}}=
 \frac{M_{B_s}}{M_{B_d}}
\xi^2 \left| \frac{V_{ts}}{V_{td}} \right|^2,
\end{equation}
where $\xi\equiv
\sqrt{B_{B_s}} f_{B_s}/(\sqrt{B_{B_d}} f_{B_d}) = 1.23 \pm 0.06$
\cite{Hashimoto:2004hn}
is a ratio of decay constants $f_{B_{s(d)}}$
and bag parameters $B_{B_{s(d)}}$ for $B_{s(d)}$ mesons.
%$\xi \equiv
%\sqrt{B_{B_s}} f_{B_s}/(\sqrt{B_{B_d}} f_{B_d}) = 1.23 \pm 0.06$
%\cite{Hashimoto:2004hn}.

%where $B_{B_{s(d)}}$ and $f_{B_{s(d)}}$ denote
%the bag parameter and decay constant of $B_{s(d)}$ meson.
%
%Since $\xi$ is measured by SU(3)
%flavor breaking irrespective of overall scale, it is more accurate
%rather than the respective decay constants and the bag parameters.
%We will use a value
%$\xi = 1.24 \pm 0.06$ \cite{Hashimoto:2004hn}.

It is convenient to parameterize the SUSY contribution by two real
parameters $C_{B_s}$ and $\phi_{B_s}$ in model-independent way as
\cite{Bona:2006sa}
\begin{equation}
C_{B_s} e^{2 i \phi_{B_s}} \equiv
\frac{M_{12}(B_s)^{}}{M_{12}(B_s)^{\rm SM}}\,,
\label{Cbs-phibs}
\end{equation}
where $M_{12}(B_s)^{} = M_{12}(B_s)^{\rm SM}+M_{12}(B_s)^{\rm SUSY}$
 denotes the off-diagonal element of $B_s$-$\bar B_s$ mass matrix.
Superscript SM (SUSY) stands for SM (SUSY) contribution.
The mass difference is given as $\Delta M_s = 2 |M_{12}(B_s)^{}|$.
%in the full Lagrangian.
%
%$M_{12}(B_s)^{\rm SM}+M_{12}(B_s)^{\rm NP}$.
%
%The two real parameters $C_{B_s}$ and $\phi_{B_s}$ are constrained
%by experiments.
%

We now discuss the SUSY contributions in $B_s$-$\bar B_s$ mixing.
When the flavor degeneracy is assumed in the MSSM, %at cut-off scale,
%the quark Yukawa matrices are the only sources of
%flavor violation. In this case,
the chargino diagram dominates the SUSY contributions of $M_{12}(B_{s,d})$.
In this case, $\phi_{B_s} \simeq 0$ in Eq.(\ref{Cbs-phibs}),
and
%there is a simple relation:
%$\Delta M_s^{\rm SUSY}/\Delta M_s^{\rm SM} \simeq
%\Delta M_d^{\rm SUSY}/\Delta M_d^{\rm SM}$.
%Therefore,
%in MSSM with flavor degeneracy,
the ratio of mass differences in the MSSM is almost same as in the
SM.
 In the general parameter space for the soft SUSY breaking terms,
the gluino box diagram dominates the SUSY contribution of
$M_{12}(B_s)$.
The gluino ($\tilde g$) contribution can be written naively in the
following mass insertion form
\begin{equation}
\frac{M_{12}^{\tilde g}}{M_{12}^{\rm SM}}
%\Delta M_s^{\tilde g}/\Delta M_s^{\rm SM}
\simeq
a\, [(\delta_{LL}^d)_{32}^2+ (\delta_{RR}^d)_{32}^2]
- b \, (\delta_{LL}^d)_{32} (\delta_{RR}^d)_{32},
\label{gluino-contribution}
\end{equation}
where  $a$ and $b$ depend on squark and gluino masses, and
$\delta_{LL,RR}^d = (M^2_{\tilde d})_{LL,RR}/\tilde m^2$ ($\tilde m$
is an averaged squark mass). The matrix $M^2_{\tilde d}$ is a
down-type squark mass matrix $ (\tilde Q, \tilde D^{c\dagger})
M^2_{\tilde d} (\tilde Q^\dagger, \tilde D^c)^{\rm T} $ in the basis
where down-type quark mass matrix is real (positive) diagonal.
When squark and gluino masses are less than 1 TeV, $a \sim O(1)$ and
$b \sim O(100)$. We also have contributions from $\delta_{LR}^d$,
but we neglect them  since they are suppressed by $(m_b/m_{\rm
SUSY})^2$.
It is worth noting that the SO(10) boundary condition gives much
larger SUSY contribution of $M_{12}(B_s)$ compared to the SU(5) case
since both off-diagonal elements for $LL$ and $RR$ are large and $b
\gg a$ in the formula, Eq.(\ref{gluino-contribution}).

%Our purpose is to see the
%correlation of $B_s$-$\bar B_s$ mixing and Br($\tau \rightarrow
%\mu\gamma$).
For the calculation of observables at weak scale, we
need to use the basis where $Y_{d,e}$ are real (positive) diagonal matrix.
In this basis, the boundary condition is given as $M^2_{\tilde Q} =
P_d M^2_{\tilde D^c} P_d^\dagger$, $M^2_{\tilde L} \simeq P_e
M^2_{\tilde E^c} P_e^\dagger$.
%The decay amplitudes, $A_L (\tau_L
%\rightarrow \mu_R \gamma$) and $A_R (\tau_R \rightarrow \mu_L
%\gamma$), are dominated by the neutralino-selectron and the
%chargino-sneutrino loop diagrams
% respectively,
%and Br($\tau \rightarrow \mu\gamma) \propto |A_L|^2 + |A_R|^2$.
%Therefore, the %2nd and 3rd generation
%relative phase between second and third generation  for $P_e$ is
%not very important for $\tau \rightarrow \mu\gamma$.
%
Since the decay width is proportional to the squared absolute values
of decay amplitudes, the phase of $\delta$ in the SUSY breaking mass
matrix at the  GUT scale and the phase of $P_e$ in the Yukawa
couplings are less important for Br($\tau \rightarrow \mu\gamma$).
On the other hand,
the phases of $\delta$ and $P_d$ are important for $M_{12}(B_s)^{\rm SUSY}$.
Due to those phases, the argument of $M_{12}(B_s)^{\rm SUSY}$
can be completely free.

%For example, in symmetric SO(10) case,
%
%the relative phase for $P_d$ is important for $M_{12}(B_s)$
%in the SO(10) case
%as
%observed from the term $- b\, (\delta_{LL}^d)_{32}
%(\delta_{RR}^d)_{32}$ in  the gluino contribution  in
%Eq.(\ref{gluino-contribution}).
%%The gluino contribution is not affected by the $\delta$'s phase.
%The phase of $\delta$'s does not contribute  since we
%have
%$(\delta_{LL}^d)_{32} \propto \delta^*$ and
%$(\delta_{RR}^d)_{32} \propto \delta\, e^{i(\phi_s-\phi_b)}$
%where $P_d = {\rm diag} (e^{i \phi_d}, e^{i
%\phi_s}, e^{i \phi_b})$.
%%$M_{\tilde d}^2{}_{LL}= M^2_{\tilde Q} + Y_d Y_d^\dagger v_d^2
%%+ (D$-term) and $M_{\tilde d}^2{}_{RR}= M^{2 \ \rm T}_{\tilde D^c} +
%%Y_d^\dagger Y_d v_d^2 + (D$-term).
%(Due to the RGE effect,
%$\delta$'s phase can have a small effect to the $\Delta M_s$.) When $e^{i
%(\phi_s - \phi_b)} \simeq +1$ , the gluino contribution is negative and the
%$\Delta M_s$ prediction is smaller than the
%SM value. %which is preferable direction by experiment.
%%The relative phase $\phi_s-\phi_b$ is an important parameter
%%to fit fermion masses and mixings in minimal SO(10) model~\cite{Babu:1992ia}.
%% the squark mass matrix is given as
%%
%%

%We note that the phase of $M_{12}(B_s)$ is completely free
%by $\delta$'s phase for SU(5)

\begin{figure}[t]
 \center
 \includegraphics[viewport = 0 4 280 208,width=8.5cm]{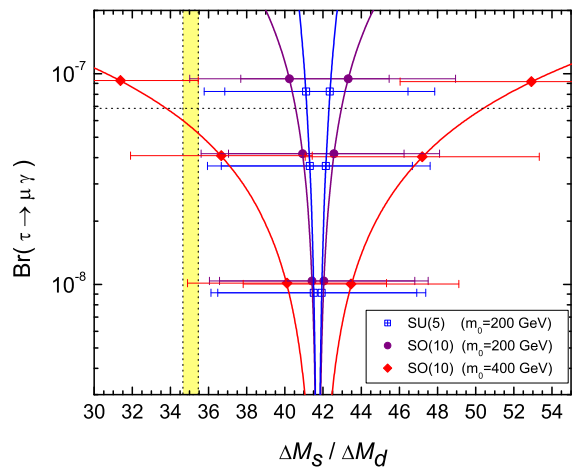}
 \caption{Minimal and maximal values for ratio of mass differences versus
          Br($\tau\rightarrow\mu\gamma$) under SU(5) and SO(10) boundary
          conditions. %with $m_0=200,400$ GeV.
 %        We use $\xi = 1.23 \pm 0.06$ and $|V_{td}/V_{ts}|=0.192\pm 0.009$ \cite{Bona:2006sa}.
          We show $|\delta| = 0.05, 0.1, 0.15$ points.
          The dotted lines show  90\% CL region of the experimental data.}
\end{figure}

In Fig.1, we plot the maximal and minimal values of the ratio of
mass differences versus branching ratio of
$\tau\rightarrow\mu\gamma$ in the case of $\tan\beta = 10$ where
$\tan\beta$ is a ratio of up- and down-type Higgs vacuum expectation
values. In the plot, we use $M_{1/2} = 300$ GeV and $A_0 =0$ for the
universal gaugino mass and the universal trilinear scalar coupling
coefficient at GUT scale. The universal scalar mass at the GUT scale
is $m_0 =200$ for the SU(5) plot, and $m_0 = 200, 400$ GeV for the
SO(10) plots.
%
%The Br($\tau\rightarrow\mu\gamma$) is larger in SO(10) compared to the SU(5) case
%due to  the neutralino contribution in $\tau_L \rightarrow \mu_R \gamma$
%for SO(10) boundary condition.
%
%Therefore, the SU(5) plot is parabolic, while SO(10) plot
%is not parabolic.
%Because of the gaugino loops in RGEs, the
%diagonal elements of SUSY breaking scalar masses become bigger at
%the weak scale especially for squarks, while the off-diagonal
%elements do not change much.
%Therefore the $\delta_{LL,RR}^{d,e}$
%values are different at the weak scale.
%
%For the plot,
%
We use $|V_{td}/V_{ts}|=0.192\pm 0.009$
which is obtained by the global CKM parameter fit without using
experimental data for $\Delta M_s$~\cite{Bona:2006sa}.
The ratio of mass differences is proportional to $\xi^2/|V_{td}|^2$.
%
%We show the plot using the
%symmetric boundary condition Eq.(\ref{symbc}).
%The plots do not
%change much if we choose the hermitian boundary condition
%Eq.(\ref{herbc}).
%
The Br($\tau\rightarrow\mu\gamma$) is almost proportional to
$\tan^2\beta$, while $\Delta M_s/\Delta M_d$ does not depend on
$\tan\beta$ much.
%
%It is easy to see Br($\tau\rightarrow\mu\gamma)
%\propto \delta^2$ and $|\Delta M_s/\Delta M_d - (\Delta M_s/\Delta
%M_d)^{\rm SM}| \propto \delta$  in SU(5) and $\propto \delta^2$ in SO(10).
%For $|\delta|\alt 0.05$, the plots of SU(5) and SO(10) are almost same
%since $(\delta_{LL}^d)_{32}$ is generated by RGE at weak scale.
%
We find that Br($\tau \to \mu\gamma$) does not have much dependence on $m_0$
(for $m_0 \alt 500$ GeV),
while the SUSY contribution of $\Delta M_s$ depends on $m_0$.
%This
%is a consequence of the fact that squark masses (diagonal elements
%of squark mass matrices) have large  gluino loop contribution rather
%than the $m_0$ contribution, while the weak gaugino loop
%contributions are less to the slepton masses and the off-diagonal
%elements of the scalar masses are not affected by the gaugino loops.

%The ratios, using the SU(5) boundary conditions in Fig.1 plots, may
%be consistent with experimental value. On the other hand, the region
%under the SO(10) boundary condition is much wider than the
%experimental allowed region. However, one can always find an
%experimentally allowed solution for the ratio of mass differences
%choosing phases in the boundary condition.

\begin{figure}[t]
 \center
 \includegraphics[viewport = 0 4 280 208,width=8.5cm]{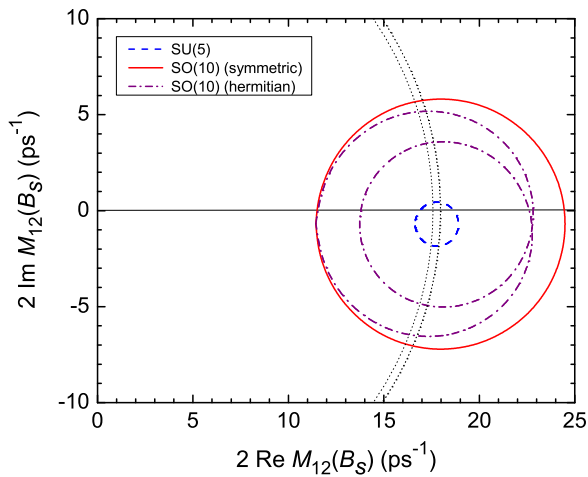}
 \caption{Re-Im plot for $2 M_{12}(B_s)$ when Br($\tau\to\mu\gamma$) saturates
          experimental bound.
          We use $\sqrt{B_{B_s}}f_{B_s} = 262$ MeV \cite{Hashimoto:2004hn}.
          The dotted lines show 90\% CL region of the experimental data.}
\end{figure}

In order to illustrate that any phase is possible for $M_{12}(B_s)^{\rm SUSY}$,
we plot the real and the imaginary part of $M_{12}(B_s)$
in the case where Br($\tau\to\mu\gamma) = 6.8 \times 10^{-8}$
\cite{Aubert:2005ye} for $\tan\beta = 10$ in Fig.2.
We use the same values for $M_{1/2}$ and $A_0$ as in Fig.1
and $m_0 = 500$ GeV.
%
%For lighter squarks and gluinos, the circle can be larger.
%In the plot, we use the Particle Data Group convention of CKM matrix \cite{Eidelman:2004wy}.
%
Using SU(5) and SO(10) (hermitian) boundary mass values, we vary the
phase of $\delta$. With SO(10) symmetric boundary conditions, we fix
$\delta$ to be real (positive) and vary the phase in the diagonal
phase matrix $P_d$. In the case of SO(10) with hermitian boundary
conditions, the plot is a double-circle due to the gluino
contribution as shown in Eq.(\ref{gluino-contribution}). Due to the
RGE effect, the double-circle does not overlap completely.
We note that even if
the radius of circle becomes large, $\Delta M_s = 2
|M_{12}(B_s)|$ has experimentally allowed solutions,
as long as $|M_{12}(B_s)^{\rm SUSY}| \alt 2 |M_{12}(B_s)^{\rm SM}|$,
though one needs
to adjust
the phases in boundary conditions. %if the radius is large.
It is worth emphasizing that the phases of $M_{12} (B_s)$ are large in such
solutions.
%
%The phase ($\sin2 {\rm Arg} M_{12}(B_s)$) can be measured
%by CP asymmetry of the decay $B_s \rightarrow J/\psi \phi$. The new
%phase also can be measured in the decay $B_s \rightarrow l^{-}X$.

\begin{figure}[t]
 \center
 \includegraphics[viewport = 0 0 280 202,width=8.5cm]{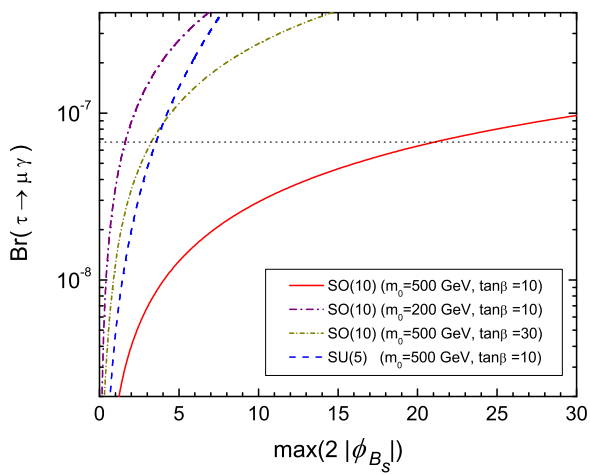}
 \caption{${\rm max}(2\,|\phi_{B_s}|)$ (degree) versus Br($\tau \rightarrow \mu\gamma$).
          %We use $m_0 = 200$ GeV and $M_{1/2} = 300$ GeV.
          %In SM, Arg $M_{12}(B_s)\simeq -2.2^{\rm o}$.
          }
\end{figure}

In order to see the maximum allowed phase in the $B_s$-$\bar B_s$ mixing, we
plot the maximal value of $2\,|\phi_{B_s}|$ versus
Br($\tau \rightarrow \mu\gamma$) in Fig.3. We use the same values for
$M_{1/2}$ and $A_0$ as in Fig.1. When $\delta m_0^2$ in the boundary
condition becomes large, the radius of the circle in  Fig.2 becomes
large, and $|\phi_{B_s}|$ can be large consequently. One can
approximately obtain ${\rm max}(\sin 2\,|\phi_{B_s}|)\simeq
|M_{12}^{\tilde g}/M_{12}^{\rm SM}|$. When the $|\phi_{B_s}|$ is
maximized, we find $C_{B_s}\simeq \cos {\rm max}(2\,|\phi_{B_s}|)$.
The model independent constraints for $C_{B_s}$ and $\phi_{B_s}$ are
$C_{B_s} = 0.97 \pm 0.27$ and $2\phi_{B_s} = (-4\pm 30)^{\rm o} \cup
(186\pm 30)^{\rm o}$~\cite{Bona:2006sa}. The phase $\phi_{B_s}$ can
be measured by CP asymmetry of the decay $B_s \rightarrow J/\psi
\phi$ and the semi-leptonic decay $B_s \rightarrow l^{-}X$.
The phase $\phi_{B_s}$ in the SO(10) case can be larger than  in the
SU(5) case since both $LL$ and $RR$ elements can be large in the
SO(10) case.
%since the gluino contribution is $\propto \delta^2$
%in SO(10) and $\propto \delta$ in SU(5) case.
As depicted in Fig.3, there is a chance for $2\phi_{B_s}$ to be
around $20^{\rm o}$ in the SO(10) case before the parameter space
gets ruled out by the Br($\tau\to\mu\gamma$). In order to obtain a
large phase, large $m_0$ and small $\tan\beta$ are needed, which
leads to an important implication. The Higgs mass bounds for MSSM
restricts the lower values of $\tan\beta$ and $M_{1/2}$.
%In the
%neutralino dark matter scenario
In the minimal supergravity model, the scalar mass $m_0$ is restricted to
be less than around 200 GeV (for $m_0<1$ TeV) \cite{Khotilovich:2005gb}
for $\tan\beta=10$ by the WMAP data \cite{sp}, and as a result,
the phase $|\phi_{B_s}|$ can not be very large. Interestingly, the
large $m_0$ solution ($m_0> 1$ TeV) for
 dark matter content~\cite{Battaglia:2003ab} may generate large
 $\phi_{B_s}\sim 90^{\rm o}$ which is allowed by the experimental data.
 However, the muon $g-2$ \cite{Bennett:2002jb} (using the $e^+e^-$ data)
 restricts $m_0 \alt500$ GeV at the 2 sigma level when $\tan\beta =10$.

In conclusion, we have studied the correlation between
$\tau\rightarrow \mu\gamma$ and $B_s$-$\bar B_s$ mixing using SU(5)
and SO(10) models.
The SO(10)  GUT models can have larger effects
on  $B_s$-$\bar B_s$ mixing %and the Br($\tau \rightarrow \mu\gamma$)
compared to the SU(5) boundary conditions.
We find that when Br($\tau \rightarrow \mu\gamma$) is enhanced
around the current experimental bound, $B_s$-$\bar B_s$ mixing may
also contain large contribution from SUSY. Consequently, the phase
of $M_{12}(B_s)$ is large for a SO(10) model and can be tested by
measuring CP asymmetries of the $B_s$ decay modes.
It is interesting to note that the phase of $B_s$-$\bar B_s$ mixing
can be large for smaller $\tan\beta$ and large scalar mass $m_0$ at
the GUT scale.
%

%Ambiguity of lattice calculation?
%
%In the near future, one can verify whether we have new physics
%in the flavor changing processes.


\end{document}